%% file: PUB_1lepton2010.tex
\documentclass[twocolumn,prl,aps,showpacs]{revtex4}
\usepackage{atlasphysics}
\usepackage{subfigure}
\usepackage{mathrsfs}
\usepackage{graphicx}
\usepackage{multirow}
\usepackage{rotating}
\usepackage{amsmath}

\input{susydefs}
\newcommand{\oneleptitle}{Search for supersymmetry using final states with one lepton, jets, and missing transverse momentum with the ATLAS detector in $\sqrt{s}=7\TeV$ pp collisions}

\newcommand{\hideit}[1]{{ }}

\newcommand{\btheta}{\boldsymbol\theta}


\begin{document}

\hspace*{8.6cm} \mbox{CERN-PH-EP-2011-013, Submitted to Phys. Rev. Lett.}

\vspace*{3mm}

\title{\oneleptitle}

\author{The ATLAS Collaboration}

\begin{abstract}
This Letter presents the first search for supersymmetry
in final states containing one isolated electron or muon, jets, and missing 
transverse momentum
from $\sqrt{s}=7\TeV$ proton-proton collisions at the LHC.
The data were recorded by the ATLAS experiment during 2010 and 
correspond to a total integrated luminosity of $35~\ipb$. No
excess above the standard model background expectation is
observed. Limits are set on the parameters of the minimal supergravity
framework, extending previous limits. For $A_0 = 0 \GeV$,
$\tan \beta = 3$, $\mu>0$ and for equal squark and gluino masses,
gluino masses below $700 \GeV$ are excluded at 95\% confidence level.
\end{abstract}

\maketitle

Many extensions of the standard model predict the existence of new
colored particles, such as the squarks ($\squark$) and gluinos ($\gluino$)
of supersymmetric (SUSY) theories~\cite{susy}, which could be accessible at
the LHC. The dominant SUSY production channels are squark-(anti)squark, squark-gluino, 
and gluino-gluino pair production.
Squarks and gluinos are expected to decay to quarks and gluons
and the SUSY partners of the gauge bosons (charginos, $\chinopm$, and
neutralinos, $\nino$), leading to events with energetic jets.
In R-parity conserving SUSY models~\cite{r-parity}, the lightest supersymmetric
particle (LSP) is stable and escapes detection, giving rise
to events with significant missing transverse momentum.
In decay chains with charginos ($\squarkL \to q \chinopm$, $\gluino \to
q \bar{q}' \chinopm$), chargino
decay to the LSP can produce a high-momentum lepton.
Currently, the most stringent limits on squark and gluino masses come from
the LHC~\cite{cms-0lepsusy}  and from the
Tevatron~\cite{cdfsquarksgluinos,d0squarksgluinos,d0trilepton}.

This Letter reports on a search for events with exactly one isolated
high-transverse momentum ($\pt$) electron or muon, at least three 
high-$\pt$ jets, and significant missing transverse momentum.
An exact definition of the signal region will be given elsewhere in
this Letter.
From an experimental point of view, the requirement of an isolated
high-$\pt$ lepton suppresses the QCD multijet background  and facilitates
triggering on interesting events. 
In addition to the signal region, three control regions
are considered for the most important standard model backgrounds. A combined
fit to the observed number of events in these four regions, together
with an independent estimate of jets misidentified as leptons
in QCD multijet events, is used to search for an
excess of events in the signal region. 

The analysis is sensitive to any new physics leading to such an excess,
and is not optimized for any particular model of SUSY.
The results are interpreted within the MSUGRA/CMSSM (minimal
supergravity/constrained minimal supersymmetric standard model) 
framework~\cite{msugra,cmmsm}
in terms of limits on the universal scalar and gaugino mass parameters 
$m_0$ and $m_{1/2}$. These are presented for fixed values of the universal 
trilinear coupling parameter 
$A_0 = 0 \GeV$, ratio of the vacuum expectation values of the two Higgs
doublets $\tan \beta = 3$, and Higgs mixing parameter $\mu > 0$, in order to
facilitate comparison with previous results.

The ATLAS detector~\cite{Aad:2008zzm} is a multipurpose particle
physics apparatus with a forward-backward symmetric cylindrical
geometry and near 4$\pi$ coverage in solid angle~\cite{atlascoordinates}.
The inner tracking detector (ID) consists of a
silicon pixel detector, a silicon microstrip detector (SCT), and
a transition radiation tracker (TRT). 
The ID is surrounded by a thin superconducting
solenoid providing a 2 T magnetic field, and by
high-granularity liquid-argon (LAr) sampling electromagnetic calorimeters.
An iron-scintillator tile calorimeter provides hadronic coverage 
in the central rapidity range.
The end-cap and forward regions 
are instrumented with LAr calorimetry
for both electromagnetic and hadronic measurements.
The muon spectrometer (MS) surrounds the calorimeters  and consists 
of three large superconducting toroids, a system 
of precision tracking chambers,
and detectors for triggering.

The data used in this analysis were recorded in 2010 
at the LHC at a center-of-mass energy of 7 \TeV. 
Application of beam, detector, and data-quality requirements results in
a total integrated luminosity of $35$ pb$^{-1}$, with an estimated
uncertainty of 11\%~\cite{atlas-lumi}.
The data have been selected with single lepton ($e$ or $\mu$) triggers. The detailed
trigger requirements vary throughout the data-taking period, but the
thresholds are always low enough to ensure that leptons with $\pt > 20 \GeV$ lie
in the efficiency plateau.

Fully simulated Monte Carlo event samples are used to develop and validate the 
analysis procedure,
compute detector acceptance and reconstruction efficiency, and aid
in the background determination.
Samples of events for background processes are generated as described
in detail in Ref.~\cite{atlas-topxsec}. For the major backgrounds, top quark pair and
$W$+jets production, MC@NLO~\cite{mcatnlo} v3.41 and ALPGEN~\cite{alpgen} v2.13 are used.
Further samples include QCD multijet events, single top production, 
diboson production, and Drell-Yan dilepton events.

Monte Carlo signal events are generated with Herwig++~\cite{herwigpp} v2.4.2. 
The SUSY particle spectra and decay modes are calculated with ISAJET~\cite{isajet}
v7.75. The SUSY samples are normalized using next-to-leading order (NLO) cross sections
as determined by Prospino~\cite{prospino} v2.1.
All signal and background samples are produced using the ATLAS MC09 parameter 
tune~\cite{mc09tunes} and a GEANT4 based~\cite{geant4} detector 
simulation~\cite{atlassimulation}.

Criteria for electron and muon identification closely follow those
described in Ref.~\cite{atlas-winclusive}.
Electrons in the signal region are required to pass the ``tight'' selection
criteria, with $\pt > 20~\GeV \,$ and $|\eta| <~2.47$.
Events are always vetoed if a ``medium'' electron is found in the electromagnetic
calorimeter transition region, $1.37 < |\eta| <~1.52$.

Muons are required to be identified either in both ID
and MS systems (combined muons)  or as a match
between an extrapolated ID track and one or more segments in the MS.
The ID track is required to have at least one pixel hit, more than five SCT hits, 
and a number of TRT hits that varies with $\eta$.
For combined muons, a good match between ID and MS tracks is required, and
the $\pt$ values measured by these two systems
must be compatible within the resolution. The summed $\pt$ of other ID tracks
within a distance $\Delta R = \sqrt{(\Delta \eta)^2 + (\Delta \phi)^2} < 0.2$ 
around the muon track is required to be less than $1.8$ \GeV.
Only muons with $\pt > 20 \GeV$ and $|\eta| < 2.4$ are considered.
For the final selection, the distance between the $z$ coordinate of the 
primary vertex and that of the extrapolated muon track at the point
of closest approach to the primary vertex
must be less than $10$ mm.

Jets are reconstructed using the anti-$k_t$ jet clustering 
algorithm~\cite{Cacciari:2008gp} with
a radius parameter $R = 0.4$. The inputs to this algorithm are clusters of
calorimeter cells seeded by cells with energy significantly above the measured
noise. Jets are constructed by performing a four-vector sum over
these clusters, treating each cluster as an $(E,\vec{p})$ four-vector with zero mass.
Jets are corrected for calorimeter non-compensation, upstream material
and other effects using $\pt$- and $\eta$-dependent calibration
factors obtained from Monte Carlo and validated with extensive
test-beam and collision-data studies~\cite{atlascalibration}. 
Only jets with $\pt > 20~\GeV \,$
and $|\eta| < 2.5$ are considered.
If a jet and a ``medium'' electron are both identified within a distance
$\Delta R < 0.2$ of each other, the jet is discarded.
Furthermore, identified ``medium'' electrons or muons are only considered if they 
satisfy $\Delta R > 0.4$ with respect to the closest remaining jet.
Events are discarded if they contain any jet failing
basic quality selection criteria, which reject detector noise and 
non-collision backgrounds~\cite{atlas-jet-clean}.

The calculation of the missing transverse momentum, $\met$, is based on the
modulus of the vectorial sum of the $\pt$ of the reconstructed objects (jets
with $\pt > 20 \GeV$, but over the full calorimeter coverage $|\eta| < 4.9$, 
and the selected lepton), any additional non--isolated muons and the
calorimeter clusters not belonging to reconstructed objects.

Events are required to have at least one reconstructed primary
vertex with at least five associated tracks.
The selection criteria for signal and control regions are based on
Monte Carlo studies prior to examining the data.
The signal region is defined as follows.
At least one identified electron or muon
with $\pt > 20~\GeV \,$ is required. Events are rejected if they
contain a second identified lepton with $\pt > 20~\GeV$,
because they are the subject of a future analysis.
At least three jets with $\pt > 30~\GeV \,$ are required, the leading one
of which must have $\pt > 60~\GeV$. In order to reduce the
background of events with fake $\met$ from
mismeasured jets, the missing transverse momentum 
vector $\vec{E}_{\mathrm{T}}^{\mathrm{miss}}$ is required
not to point in the direction of any of the three leading jets:
$\Delta\phi(\textrm{jet}_{i},\vec{E}_{\mathrm{T}}^{\mathrm{miss}}) > 0.2 \, (i=1,2,3)$.
The transverse mass between the lepton and
the missing transverse momentum vector,
$\mt\ =\sqrt{2\cdot p_{\mathrm{T}}^\ell\cdot\met\cdot(1 - \cos(\Delta\phi(\ell,\met)))}$,
is required to
be larger than $100 \GeV$. $\met$ must
exceed $125 \GeV$ and must satisfy $\met > 0.25 \,  \meff$,
where the effective mass $\meff$ is the scalar sum of the
$\pt$ of the three leading jets, the $\pt$ of the 
lepton, and $\met$. Finally, a cut is applied on the
effective mass: $\meff > 500$ \GeV. 
The efficiency for the SUSY signal
in the MSUGRA/CMSSM model defined earlier varies between 0.01\% for $m_{1/2} = 100 \GeV$
and 4\% for $m_{1/2} = 350 \GeV$, with a smaller dependence on $m_0$, for
the electron channel and the muon channel separately. The inefficiency is
dominated by the leptonic branching fractions in the SUSY signal.

Backgrounds from several standard model processes could contaminate
the signal region.
Top quark pair production and $W$+jets production backgrounds are
estimated from a combined fit to the number of observed events in
three control regions, using Monte Carlo simulations to derive the
background in the signal region from the control regions.
The background determination of
QCD multijet production with a jet misidentified as an isolated lepton
is purely data driven.
Remaining backgrounds from other sources are estimated with simulations.

The three control regions have identical
lepton and jet selection criteria as the signal region.
The top control region
is defined by a window in the two-dimensional plane of
$30 \GeV < \met < 80 \GeV$ and $40 \GeV < \mt < 80 \GeV$  and by
requiring that at least one of the three leading
jets is tagged as a $b$-quark jet. For
the $b$-tagging, the secondary vertex algorithm SV0~\cite{sv0} is used, 
which, for $\pt = 60 \GeV$ jets, provides an efficiency of 
50\% for $b$-quark jets and a
mistag rate of 0.5\% for light-quark jets.
The $W$ control region is defined by the same window in the
$\met - \mt$ plane, but with the requirement
that none of the three hardest jets is $b$-tagged.
The QCD multijet control region is defined by demanding low
missing transverse momentum, $\met < 40 \GeV$, and low transverse
mass, $\mt < 40 \GeV$. This QCD control region is only used to estimate 
the QCD multijet background contribution to other background regions but not
to the signal region.
Instead, the electron and muon identification criteria are relaxed,
obtaining a ``loose'' control sample that is dominated by QCD jets.
A loose-tight matrix method, in close analogy to that described 
in Ref.~\cite{atlas-topxsec}, is then used to estimate the number of 
QCD multijet events with fake leptons in the signal region after
final selection criteria:
$0.0^{+0.5}_{-0.0}$ in the muon
channel and $0.0^{+0.3}_{-0.0}$ in the electron channel.

\begin{figure}
  \centering
  \includegraphics[width=0.9\columnwidth]{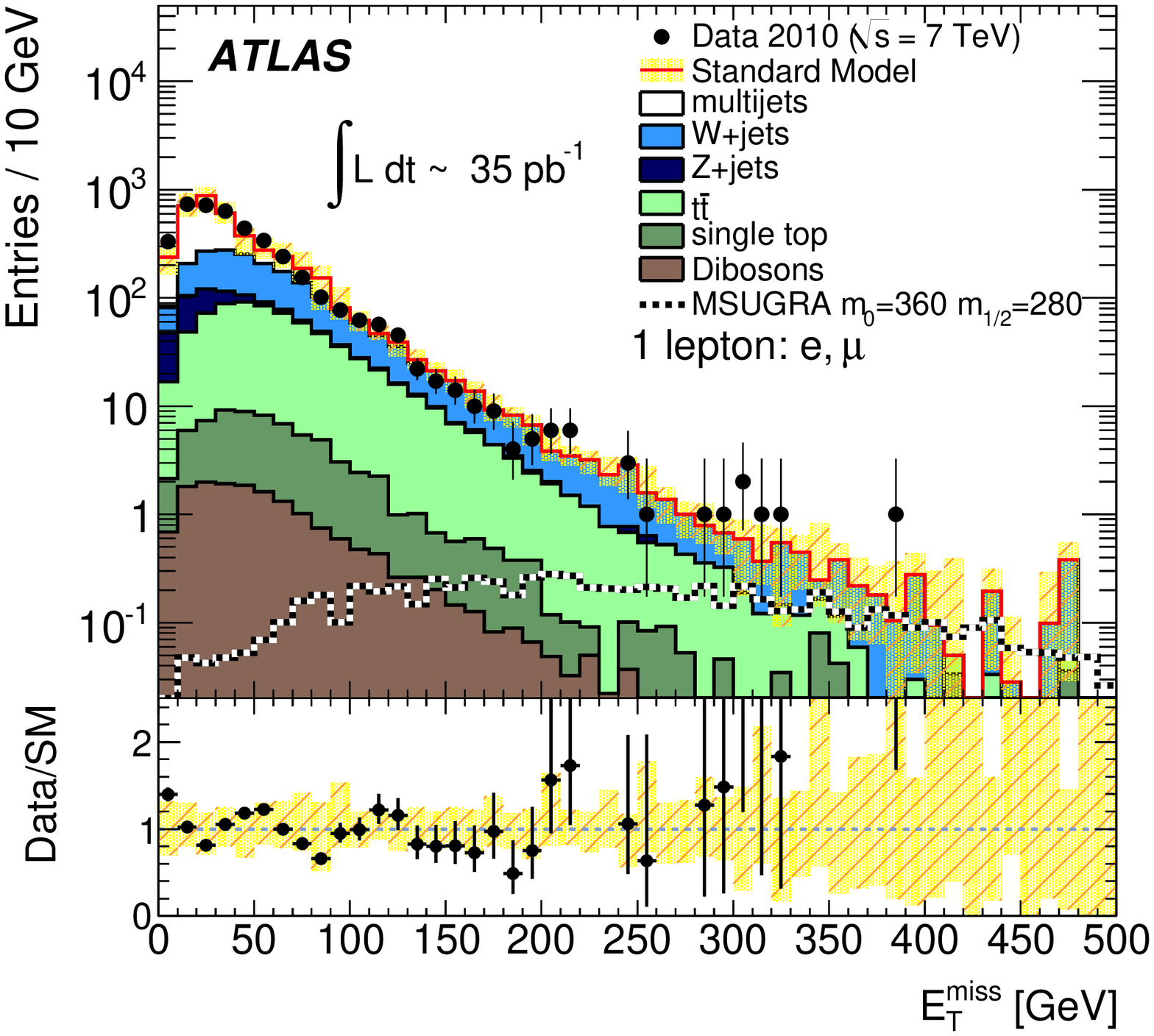}
  \includegraphics[width=0.9\columnwidth]{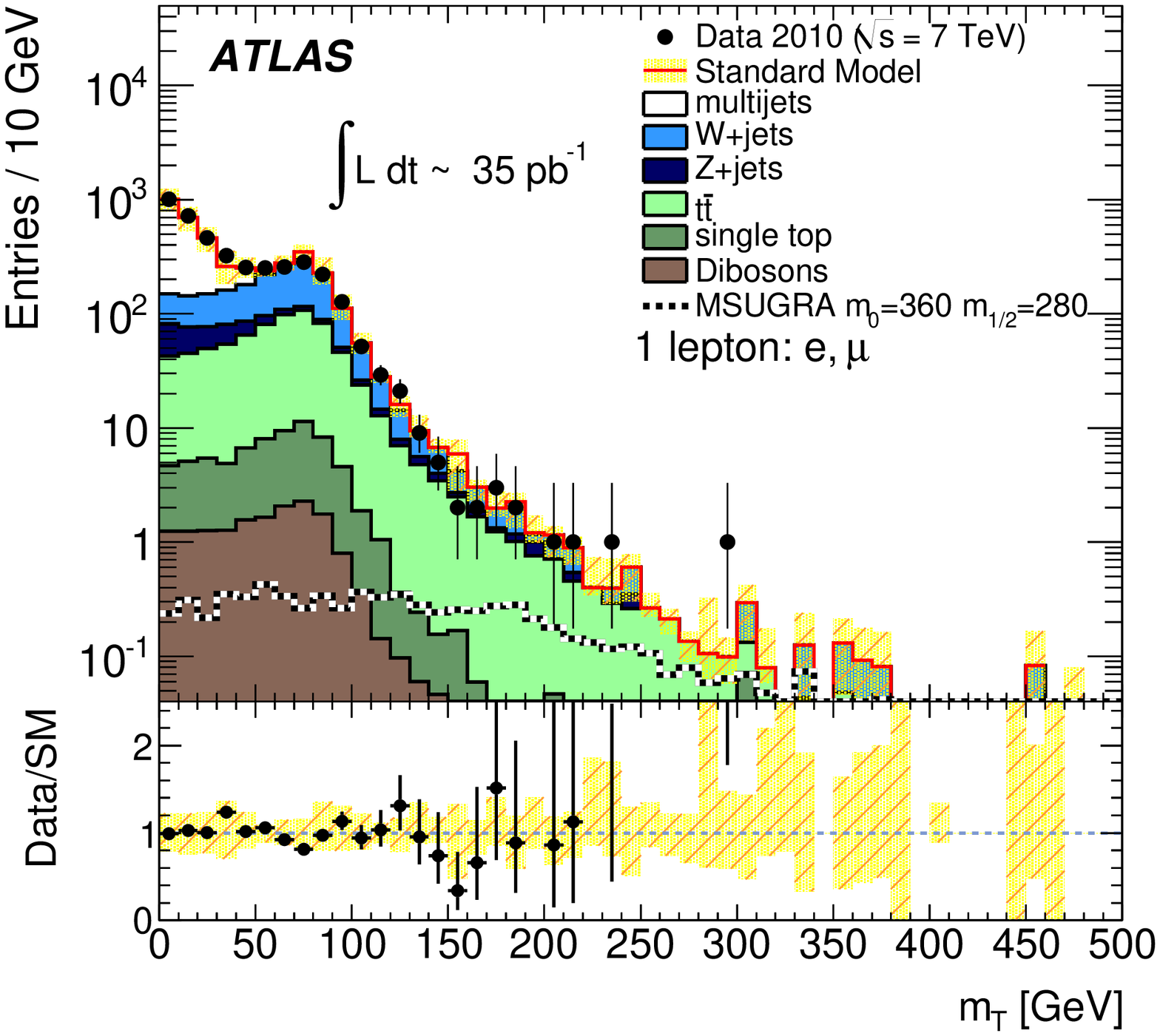}
  \includegraphics[width=0.9\columnwidth]{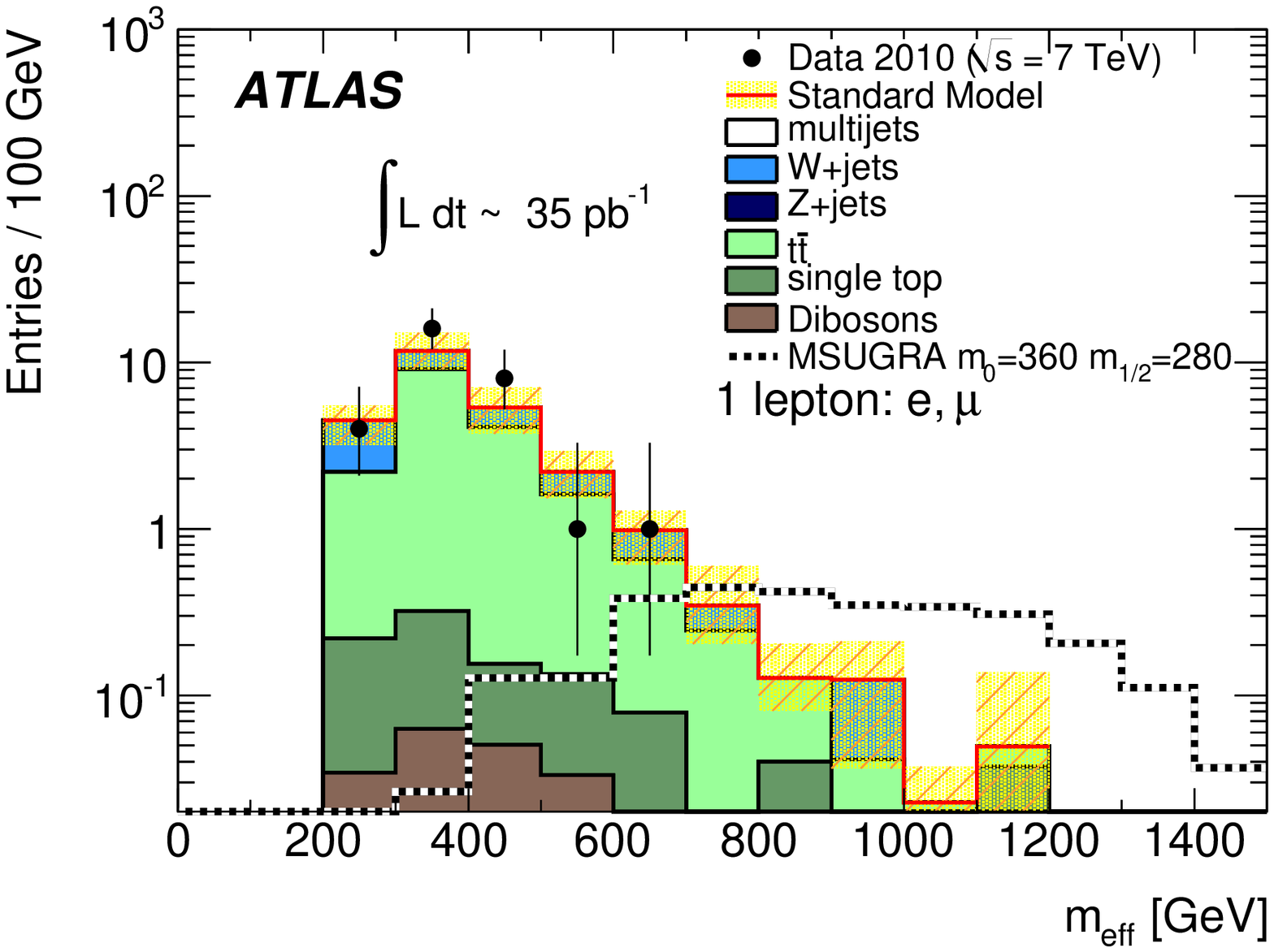}
  \caption{Top: $\met$ distribution after lepton and jet selection. Center: $\mt$
distribution after lepton and jet selection.
Bottom:
Effective mass distribution after final selection criteria except
for the cut on the effective mass itself.
All plots are made for the electron and muon channel combined.
Yellow bands indicate the uncertainty on the Monte Carlo prediction
from finite Monte Carlo statistics and from the jet energy scale uncertainty.
}
  \label{fig:controlDistributions1}
\end{figure}

Data are compared to expectations in Figure~\ref{fig:controlDistributions1}.
The standard model backgrounds in the figure are normalized to the theoretical cross sections,
except for the multijet background which is normalized to data in the
QCD multijet control region. The data are in good agreement with the standard
model expectations.
After final selection, one
event remains in the signal region in the electron channel and one
event remains in the muon channel.
Figure~\ref{fig:controlDistributions1} also shows the expected distributions for
the MSUGRA/CMSSM model point $m_0 = 360 \GeV$ and $m_{1/2} = 280 \GeV$. 

A combined fit to the number of observed events in the signal and 
control regions is performed. 
The assumption that the Monte Carlo is able to predict the backgrounds
in the signal region from the control regions is validated by checking additional
control regions at low $\mt$ and at low $\met$.
The defined control regions are not completely pure, and the combined fit
takes the expected background cross-contaminations into account.
The likelihood function of the fit
can be written as: $L({\bf n}|s,{\bf b},{\btheta}) = P_{\mathrm{S}} \,\times\,
P_{\mathrm{W}} \, \times \, P_{\mathrm{T}} \, \times \,
P_{\mathrm{Q}} \, \times \, C_{\mathrm{Syst}}$, where ${\bf n}$ represents
the number of observed events in data, $s$ is the SUSY signal to be tested,
${\bf b}$ is the background, and ${\btheta}$ represents the systematic 
uncertainties, which are treated as
nuisance parameters with a Gaussian probability density function. 
The four $P$ functions in the right hand side
are Poisson probability distributions for event counts in the defined 
signal (S) and control regions (W, T, and Q for $W$, top pair and QCD multijets
respectively),
and $C_{\mathrm{Syst}}$ represents the constraints on systematic
uncertainties, including correlations.

The dominant sources of systematic uncertainties in the background
estimates arise from Monte Carlo modeling of the shape of the
$\met$ and $\mt$ distributions in
signal and control regions. These uncertainties
are determined by variation of the Monte Carlo generator,
as well as by variations of internal generator parameters.
Finite statistics in the background control regions also contributes to
the uncertainty.
Experimental uncertainties are varied within their determined range
and are dominated by the jet energy scale 
uncertainty~\cite{atlas-inclusivejet}, $b$-tagging
uncertainties, and the uncertainty on the luminosity. 

Systematic uncertainties on the SUSY signal are estimated by variation
of the factorization and renormalization scales in Prospino 
and by including the parton density function (PDF) uncertainties
using the eigenvector sets provided by CTEQ6~\cite{cteq66}.
Uncertainties are calculated separately for the individual production processes.
Within the relevant kinematic range, typical uncertainties resulting
from scale variations are
10--16\%, whereas PDF uncertainties vary from 5\% for $\squark \squark$
production to 15--30\% for $\gluino \gluino$ production.

The result of the combined fit to signal and control 
regions, leaving the number of signal events free in the signal region
while not allowing for a signal contamination in the other regions, is shown 
in Table~\ref{tab:fitresults}. 
The observed number
of events in data is consistent with the standard model expectation.

\begin{table*}[tb]
\begin{center}
\caption{Numbers of observed events in the signal and background control
regions, as well as their estimated values from the fit (see text), for
the electron (top part) and muon (bottom part) channels. The central values of
the fitted sum of backgrounds in the control regions agree with the
observations by construction.
For comparison, nominal Monte Carlo expectations are given in parentheses for the
signal region, the top control region and the $W$ control region.}
\label{tab:fitresults}
\setlength{\tabcolsep}{0.0pc}
{\footnotesize
\begin{tabular*}{\textwidth}{@{\extracolsep{\fill}}lrrrr}
\noalign{\smallskip}\hline\noalign{\smallskip}
 {\bf Electron channel}        &  Signal region & Top region & $W$ region & QCD region   \\[-0.05cm]
\noalign{\smallskip}\hline\noalign{\smallskip}
Observed events &  $1$ & $80$ & $202$ & $1464$ \\
\noalign{\smallskip}\hline\noalign{\smallskip}
Fitted top events & $1.34 \pm 0.52$ ($1.29$) & $65 \pm 12$ ($63$) & $32 \pm 16$ ($31$) &  $40 \pm 11$\\
Fitted $W/Z$ events & $0.47 \pm 0.40$ ($0.46$) & $11.2 \pm 4.6$ ($10.2$) & $161 \pm 27$ ($146$) & $170 \pm 34$ \\
Fitted QCD events & $0.0 {}^{+0.3}_{-0.0}$ & $3.7 \pm 7.6$& $9 \pm 20$ & $1254 \pm 51$ \\
\noalign{\smallskip}\hline\noalign{\smallskip}
Fitted sum of background events & $1.81 \pm 0.75$ & $80 \pm 9$ & $202 \pm 14$& $1464 \pm 38$\\
\noalign{\smallskip}\hline\noalign{\smallskip}\noalign{\smallskip}
\end{tabular*}
\begin{tabular*}{\textwidth}{@{\extracolsep{\fill}}lrrrr}
\noalign{\smallskip}\hline\noalign{\smallskip}
 {\bf Muon channel}        &  Signal region & Top region & $W$ region & QCD region   \\[-0.05cm]
\noalign{\smallskip}\hline\noalign{\smallskip}
Observed events &  $1$ & $93$ & $165$ & $346$ \\
\noalign{\smallskip}\hline\noalign{\smallskip}
Fitted top events & $1.76 \pm 0.67$ ($1.39$) & $85 \pm 11$ ($67$) & $42 \pm 19$ ($33$) & $50 \pm 10$\\
Fitted $W/Z$ events & $0.49 \pm 0.36$ ($0.71$) & $7.7 \pm 3.3$ ($11.6$) & $120  \pm 26$ ($166$) & $71 \pm 16$ \\
Fitted QCD events & $0.0 {}^{+0.5}_{-0.0}$ & $0.3 \pm 1.2$& $3 \pm 12$    & $225 \pm 22$ \\
\noalign{\smallskip}\hline\noalign{\smallskip}
Fitted sum of background  events & $2.25 \pm 0.94$ & $93 \pm 10$ & $165 \pm 13$& $346 \pm 19$\\
\noalign{\smallskip}\hline\noalign{\smallskip}
\end{tabular*}}
\end{center}
\end{table*}

Limits are set on contributions of new physics to the signal
region. These limits are obtained from a second combined fit to the four regions,
this time allowing for a
signal in all four regions, and leaving all nuisance parameters free.
The limits are then derived from the profile likelihood ratio,
$\Lambda(s) = -2(\ln L({\bf n}|s,\hat{\hat{{\bf b}}},\hat{\hat{\btheta}}) -
\ln L({\bf n}|\hat{s},\hat{{\bf b}},\hat{\btheta}))$, where
$\hat{s}$, $\hat{{\bf b}}$ and $\hat{\btheta}$ maximize the likelihood
function and
$\hat{\hat{{\bf b}}}$ and $\hat{\hat{\btheta}}$ maximize the likelihood for
a given choice of $s$. In the fit, $s$ and $\hat{s}$ are constrained to be non-negative. 
The test statistic is $\Lambda(s)$. The exclusion $p$-values are 
obtained from this using pseudo-experiments and the limits set are
one-sided upper limits~\cite{higgscombi}.

From the fit to a model with signal events only in the signal region, 
a 95\% CL upper limit on the number of events from new physics in
the signal region can be derived. This number is $2.2$ in the electron
channel and $2.5$ in the muon channel. This corresponds to a 95\% CL upper
limit on the effective cross section for new processes in the
signal region, including the effects of
experimental acceptance and efficiency, of $0.065$ pb for the 
electron channel and $0.073$ pb for the muon channel.

\begin{figure}
  \centering
  \includegraphics[width=\columnwidth]{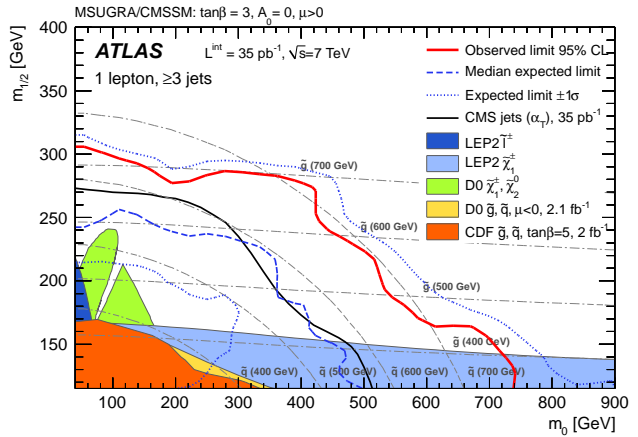}
  \caption{Observed and expected 95\% CL exclusion limits, as well as the $\pm 1 \sigma$
  variation on the expected limit, in the combined
  electron and muon channels.
  Also shown are the published limits from CMS~\cite{cms-0lepsusy},
  CDF~\cite{cdfsquarksgluinos}, and
  D0~\cite{d0squarksgluinos,d0trilepton}, and the results from the LEP 
  experiments~\cite{lepsusy}.}
  \label{fig:sugraExclusion}
\end{figure}

Within the MSUGRA/CMSSM framework, the results are interpreted as limits
in the $m_0-m_{1/2}$ plane, as shown
in Figure~\ref{fig:sugraExclusion}.
For the model considered and for equal squark and gluino masses,
gluino masses below $700 \GeV \,$ are excluded at 95\% CL.
The limits depend only moderately on $\tan \beta$.

In summary, 
the first ATLAS results on searches for supersymmetry with an isolated 
electron or muon, jets, and missing transverse momentum have been presented.
In a data sample corresponding to $35$ pb$^{-1}$, no significant deviations
from the standard model expectation are observed. Limits on the cross section
for new processes within the experimental acceptance and efficiency are set.
For a chosen set of parameters within MSUGRA/CMSSM, and for equal
squark and gluino masses, gluino masses below $700 \GeV$ are excluded
at 95\% CL. These ATLAS results exceed previous limits set by other
experiments~\cite{cms-0lepsusy,cdfsquarksgluinos,d0squarksgluinos,d0trilepton}.

\input{acknowledgement}

\bibliographystyle{atlasnote}
\bibliography{PUB_1lepton2010}

\include{atlas_authlist_final}

\end{document}

%% file: susydefs.tex
\newcommand{\meff}{\ensuremath{m_{\mathrm{eff}}}} 
\newcommand{\mt}{\ensuremath{m_\mathrm{T}}}

\def\lsim{\mathrel{\rlap{\lower4pt\hbox{\hskip1pt$\sim$}}
    \raise1pt\hbox{$<$}}}                
\def\gsim{\mathrel{\rlap{\lower4pt\hbox{\hskip1pt$\sim$}}
    \raise1pt\hbox{$>$}}}                

\makeatletter
\renewcommand{\p@subfigure}{\arabic{figure}}
\renewcommand{\thesubfigure}{\alph{subfigure}}
\renewcommand{\@thesubfigure}{(\thesubfigure)}
\makeatother

%% file: acknowledgement.tex
We wish to thank CERN for the efficient commissioning and operation of the LHC during this initial high-energy data-taking period as well as the support staff from our institutions without whom ATLAS could not be operated efficiently.

We acknowledge the support of ANPCyT, Argentina; YerPhI, Armenia; ARC, Australia; BMWF, Austria; ANAS, Azerbaijan; SSTC, Belarus; CNPq and FAPESP, Brazil; NSERC, NRC and CFI, Canada; CERN; CONICYT, Chile; CAS, MOST and NSFC, China; COLCIENCIAS, Colombia; MSMT CR, MPO CR and VSC CR, Czech Republic; DNRF, DNSRC and Lundbeck Foundation, Denmark; ARTEMIS, European Union; IN2P3-CNRS, CEA-DSM/IRFU, France; GNAS, Georgia; BMBF, DFG, HGF, MPG and AvH Foundation, Germany; GSRT, Greece; ISF, MINERVA, GIF, DIP and Benoziyo Center, Israel; INFN, Italy; MEXT and JSPS, Japan; CNRST, Morocco; FOM and NWO, Netherlands; RCN, Norway;  MNiSW, Poland; GRICES and FCT, Portugal;  MERYS (MECTS), Romania;  MES of Russia and ROSATOM, Russian Federation; JINR; MSTD, Serbia; MSSR, Slovakia; ARRS and MVZT, Slovenia; DST/NRF, South Africa; MICINN, Spain; SRC and Wallenberg Foundation, Sweden; SER,  SNSF and Cantons of Bern and Geneva, Switzerland;  NSC, Taiwan; TAEK, Turkey; STFC, the Royal Society and Leverhulme Trust, United Kingdom; DOE and NSF, United States of America.  

The crucial computing support from all WLCG partners is acknowledged gratefully, in particular from CERN and the ATLAS Tier-1 facilities at TRIUMF (Canada), NDGF (Denmark, Norway, Sweden), CC-IN2P3 (France), KIT/GridKA (Germany), INFN-CNAF (Italy), NL-T1 (Netherlands), PIC (Spain), ASGC (Taiwan), RAL (UK) and BNL (USA) and in the Tier-2 facilities worldwide.

%% file: atlas_authlist_final.tex
\begin{flushleft}
{\Large The ATLAS Collaboration}

\bigskip

G.~Aad$^{\rm 48}$,
B.~Abbott$^{\rm 112}$,
J.~Abdallah$^{\rm 11}$,
A.A.~Abdelalim$^{\rm 49}$,
A.~Abdesselam$^{\rm 119}$,
O.~Abdinov$^{\rm 10}$,
B.~Abi$^{\rm 113}$,
M.~Abolins$^{\rm 89}$,
H.~Abramowicz$^{\rm 154}$,
H.~Abreu$^{\rm 116}$
% ...
\bigskip

% ...
$^{10}$ Institute of Physics, Azerbaijan Academy of Sciences, H. Javid Avenue 33, AZ 143 Baku, Azerbaijan\\
$^{11}$ Institut de F\'isica d'Altes Energies, IFAE, Edifici Cn, Universitat Aut\`onoma  de Barcelona,  ES - 08193 Bellaterra (Barcelona), Spain\\
% ...
$^{48}$ Albert-Ludwigs-Universit\"{a}t, Fakult\"{a}t f\"{u}r Mathematik und Physik, Hermann-Herder Str. 3, D - 79104 Freiburg i.Br., Germany\\
$^{49}$ Universit\'e de Gen\`eve, Section de Physique, 24 rue Ernest Ansermet, CH - 1211 Geneve 4, Switzerland\\
% ...
$^{89}$ Michigan State University, Department of Physics and Astronomy, High Energy Physics Group, East Lansing, MI 48824-2320, United States of America\\
% ...
$^{112}$ University of Oklahoma, Homer L. Dodge Department of Physics and Astronomy, 440 West Brooks, Room 100, Norman, OK 73019-0225, United States of America\\
$^{113}$ Oklahoma State University, Department of Physics, 145 Physical Sciences Building, Stillwater, OK 74078-3072, United States of America\\
% ...
$^{116}$ LAL, Univ. Paris-Sud, IN2P3/CNRS, Orsay, France\\
% ...
$^{119}$ Oxford University, Department of Physics, Denys Wilkinson Building, Keble Road, Oxford OX1 3RH, United Kingdom\\
% ...
$^{154}$ Tel Aviv University, Raymond and Beverly Sackler School of Physics and Astronomy, Ramat Aviv, IL - Tel Aviv 69978, Israel\\
% ...
\end{flushleft}


%% file: PUB_1lepton2010.bbl
\begin{thebibliography}{10}
\bibitem{atlascoordinates}
{ ATLAS uses a right-handed coordinate system with its origin at the nominal
  interaction point (IP) in the centre of the detector and the $z$-axis
  coinciding with the axis of the beam pipe. The $x$-axis points from the IP to
  the centre of the LHC ring, and the $y$ axis points upward. Cylindrical
  coordinates $(r,\phi)$ are used in the transverse plane, $\phi$ being the
  azimuthal angle around the beam pipe. The pseudorapidity is defined in terms
  of the polar angle $\theta$ as $\eta=-\ln\tan(\theta/2)$.}
\end{thebibliography}
